\documentclass{elsart} 

\def\be{\begin{eqnarray} &&}
\def\nonu{\nonumber \\ &&}
\def\ee{\end{eqnarray}}
\def\psla{ \slash \!\!\!}
\def\Psla{ \slash \!\!\!\!}
\usepackage{graphicx}
\usepackage{epsfig}
\begin{document}

\begin{frontmatter}
\title{ Time- and Spacelike  Nucleon Electromagnetic 
Form Factors beyond Relativistic Constituent  Quark Models}
\author{J. P. B. C. de Melo$^a$, T. Frederico$^b$, E. Pace$^{c,d}$, 
S. Pisano$^{e}$ and
G. Salm\`e$^{f}$}
\address{
$^a$ Laborat\'orio de F\'\i sica Te\'orica e Computa\c c\~ao Cient\'\i fica,
Universidade Cruzeiro do Sul, 08060-700 and Instituto de F\'\i sica 
Te\'orica, 01405-900, S\~ ao Paulo, Brazil \\
$^b$ Dep. de F\'\i sica, Instituto Tecnol\'ogico de Aeron\'autica,
12.228-900 S\~ao Jos\'e dos
Campos, S\~ao Paulo, Brazil\\
$^c$ Dip. di Fisica, Universit\`a di Roma "Tor Vergata", Via della Ricerca
Scientifica 1, I-00133  Roma, Italy \\ 
$^d$ Istituto Nazionale di Fisica Nucleare, Sezione Tor Vergata, Via della Ricerca
Scientifica 1,  Roma, Italy \\
$^e$ Dip. di Fisica, Universit\`a di Roma "Sapienza", P.le A. Moro 2, 
I-00185 Roma, Italy  \\
$^f$ Istituto Nazionale di Fisica Nucleare, Sezione di Roma, P.le A. Moro 2,
 I-00185 Roma, Italy  }

\begin{abstract}
For the first time, a phenomenological analysis of the experimental 
electromagnetic form factors of
the nucleon, both in the timelike and spacelike regions, is performed 
 by taking into account the effects of  nonvalence components in the 
nucleon state, within a
light-front framework. Our model, based on 
suitable Ansatzes for the nucleon Bethe-Salpeter 
amplitude and a microscopic version of the well-known Vector Meson Dominance model,
 has only four adjusted parameters (determined by the spacelike data with 
 $\chi^2/datum \sim 1.7$), and yields a nice description of the
experimental electromagnetic form factors in the physical region in the range
 $-30 ~(GeV/c)^2~<~q^2~ < 20 ~
(GeV/c)^2$, except for the neutron one in the
timelike region. Valuable information can be gained in the timelike region on possible missing Vector 
Mesons around $q^2 \sim 4.5 ~(GeV/c)^2$ and $q^2 \sim 8.0 ~(GeV/c)^2$.
\end{abstract}

\end{frontmatter}

In recent years there has been a renewed interest in the investigation of 
nucleon electromagnetic form factors (FF), given an unexpected
discrepancy  between experimental data
for the spacelike (SL) ratio $\mu^p~G^p_E(q^2)/G^p_M(q^2)$ extracted by using: i) the Rosenbluth separation method 
(see Ref. \cite {Rosen} for
 recent measurements) and  ii) the polarization transfer technique adopted in 
 experiments carried out at TJLAB \cite{JLAB}.
Indeed in  the SL region (where the squared four-momentum transfer becomes negative,
i.e. $ q^2=  \omega^2 - |{\bf q}|^2~\le~ 0$),  data  
 obtained by the Rosenbluth separation   follow the dipole law,
 while,  surprisingly,  data from the polarization transfer technique decrease 
 faster than the dipole law
 for   $Q^2=-q^2 > 1 ~ (GeV/c)^2$.
This experimental puzzle has not yet been completely explained,
although both two-photon exchange processes \cite{KBMT}
  and higher-order radiative corrections \cite{rad} appear relevant  for 
  its solution.
 
 Furthermore,  the timelike (TL) region calls for both experimental and
 theoretical investigations (in particular for the neutron !), since the ratio, $R$,
  between {\em experimental}  
 neutron and proton form factors, beyond the threshold  $q^2= 4M^2_N$ (with
 $M_N$ the nucleon mass), turns out to be greater than one
\cite{Fenice}, while naive 
expectations from perturbative QCD (see, e.g. \cite{Karliner}) yield  
$R \sim |e_d/e_u|=0.5$.
A deeper understanding of all these experimental issues could open new windows in
the investigation of the nucleon 
internal structure,  also elucidating the role of small components in the
nucleon state, (see, e.g., \cite{miller} for  the possible influence of 
 the above mentioned SL puzzle  on the nucleon shape studies).

\begin{figure}

\includegraphics[width=12.5cm]{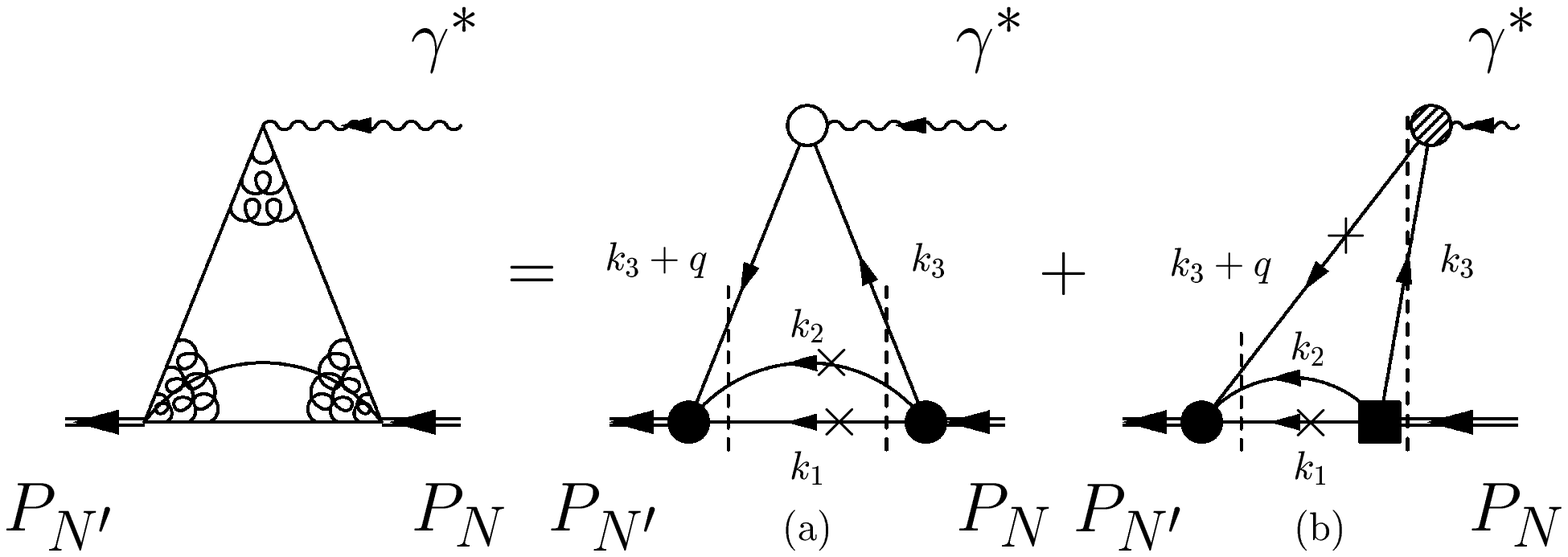}
\caption{\label{slfey}
Diagrams contributing to the SL nucleon FF: (a) valence (triangle)
contribution with $0 \le k_{i}^+ \le P^+_{N}$ (i = 1,2,3) and $0 
\le k_{3}^+ + q^+ \le P'^+_{N}$; 
(b) nonvalence, pair-production
contribution with $0 > k_3^+ \ge - q^+$. A cross on a quark line indicates a 
quark  
on the $k^-$-shell. 
Solid circles and  solid square represent  valence
and NV vertex functions, respectively;
open and  shaded circles are bare and dressed
  quark-photon vertexes, respectively.}
\vspace{.2cm}
\includegraphics[width=12.5cm]{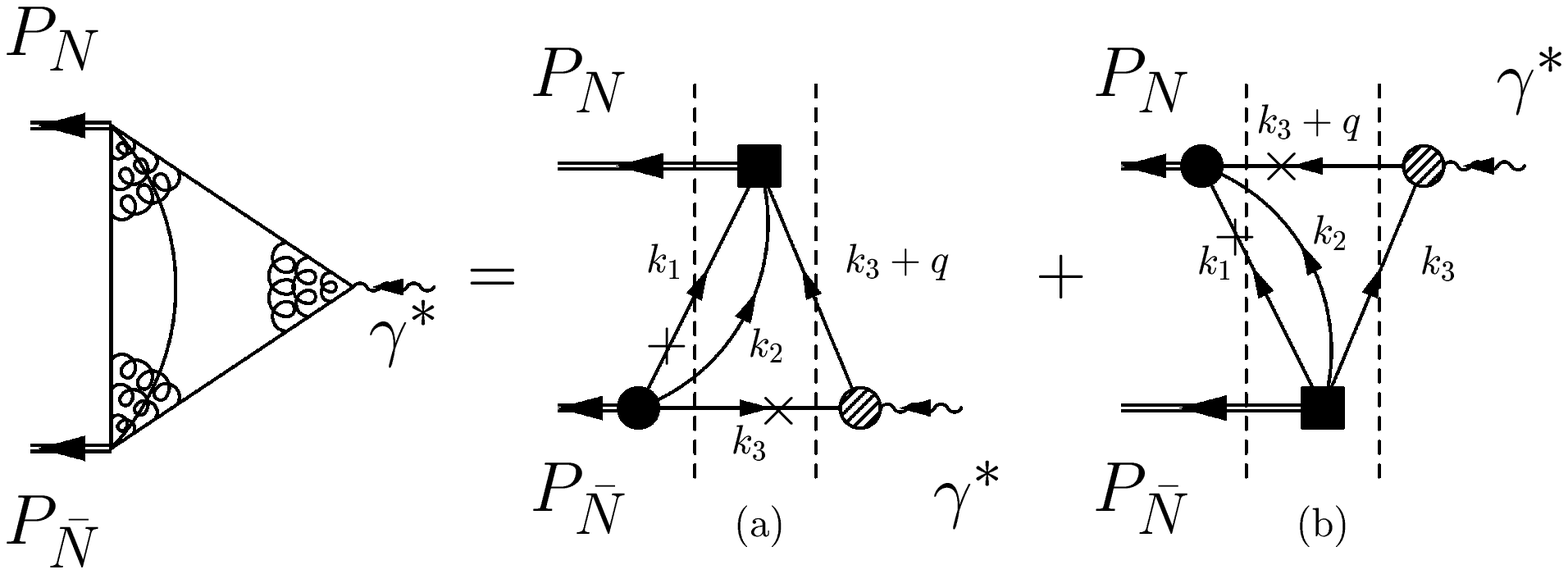}
\caption{\label{tlfey}
Diagrams contributing to the TL nucleon FF: 
(a) $P^+_{N} < k_3^+ + q^+ < q^+ $; 
(b) $0\le k_3^+ + q^+ \le P^+_{N}$.  Symbols
 as in Fig. \ref{slfey}.}
 \end{figure}

Within the light-front dynamics \cite{brodsky,karma}, we successfully 
reproduced the pion experimental FF 
in the interval $ - 10~(GeV/c)^2 \le q^2 \le  10~(GeV/c)^2$ \cite{DFPS},
 namely both in the SL and TL regions, by introducing
components of the pion state beyond the valence one.
Aim of this letter is the generalization of our approach to the nucleon,  
presenting for the first time within the light-front dynamics
a unified, direct calculation of both SL and TL nucleon FF (see, also  
\cite{MFPPS} for preliminary results with a slightly different model). 
 In particular, the  role of the
 contribution due to the $q\bar{q}$-pair, created
 by the incoming virtual photon, turns out to be essential as in the pion case. Our approach
 shares many ingredients with the model of Ref. \cite{nua}, but  it exhibits distinct
 features, like i) the choice of a reference frame with the plus component of
 the momentum transfer $q^+=q^0+q_z\ne 0$ \cite{LPS}, allowing a
 unified analysis of
 SL  and TL  regions (see Figs. \ref{slfey} and \ref{tlfey} for a diagrammatic
 illustration), 
 and ii) the gauge-invariant dressing of the quark-photon vertex through a
 microscopic   Vector Meson Model (VMD) \cite{DFPS}.

 Nucleon FF, that enter in the {\em macroscopic} description of the em current operator, 
 $I^\mu(q^2)$, are calculated in a reference frame
 where ${\bf q}_{\perp}= {\bf P_N}_{\perp} = 0$ and $q^+ = \sqrt{| q^2 |}$.
 In the SL region, 
 where  $q^\mu={P'}_N^\mu~ -~P_{ N}^\mu$ (with $P_N$  and $P_N^\prime$ 
the initial and final  nucleon four-momenta, respectively) 
 and $q^2 \le0$,  
 the nucleon Sachs FF are given by  
\be
G_{E}^N(q^2)  
 ={1 \over 2} 
Tr \left \{{\Psla P_N^\prime +M_N\over 2 M_N}~ I^{+}(q^2)~{\Psla P_N +M_N\over 2 M_N}
~ \gamma^{+}\right\} \nonu G_{M}^N(q^2)  
 = \eta
Tr \left \{{\Psla P_N^\prime +M_N\over 2 M_N}~ I_x(q^2)~{\Psla P_N +M_N\over 2 M_N}
~ \gamma_x\right\}
\label{ffi}
\ee
where $\eta=-2 M_N^2/q^2$.
The expressions for the TL form factors corresponding to Eq. (\ref{ffi})
  can be easily obtained by changing $P_N$ with
 $-P_{\bar N}$ and $P'_N$ with $P_N$. In our approach, for the SL kinematics,
 the matrix elements of the  nucleon current (see 
 \cite{MFPPS}) 
are approximated {\em microscopically} in impulse approximation
by the Mandelstam formula \cite{mandel} as follows (see also \cite{nua})
\be
\bar U^{\sigma'}_{N'} ~I^\mu(q^2)~ U^\sigma_N = 
3 ~ N_c  \int {d^4k_1 \over (2\pi)^4}\int {d^4k_2 \over (2\pi)^4} 
  \sum
\left \{\bar \Phi^{\sigma'}(k_1,k_2,k'_3,P_N')~S^{-1}(k_1)~\right.
\times \nonu \left.S^{-1}(k_2)~ {\mathcal I}^\mu(k_3,q)~
 ~\Phi^\sigma(k_1,k_2,k_3,P_N)\right \}
\label{spc1}
\ee
 where $U^\sigma_N$  is the nucleon  Dirac spinor, the factor 3 comes from the 
symmetry of our problem, $N_c$ is the number of colors,  $\Phi^\sigma(k_1,k_2,k_3,P_N)$
the nucleon Bethe-Salpeter amplitude (BSA), $k_i$ the i-th constituent 
quark (CQ) four-momentum,
$k'_3 = k_3 +q$, $P_N = k_1+k_2+k_3$ and  $P'_N=k_1+k_2+k'_3$ . 
In Eq. (\ref{spc1}), $ \sum$ implies a 
sum over isospin and spinor indexes, $S(k_i)$ is
the Dirac  propagator  of a CQ (with a chosen constituent mass $m=200~MeV$) and 
${\mathcal I}^\mu(k_3,q)$ is
 the  quark-photon vertex,  obtained by dressing a
pointlike quark (see below). An  expression analogous to Eq. (\ref{spc1}) holds for the TL
region (see \cite{MFPPS2}).

The nucleon BSA
must have a  Dirac structure, that has been devised exploiting a $qqq-N$
  effective Lagrangian which couples a scalar-isoscalar quark-pair plus a quark to the nucleon,
   as suggested in Ref. \cite{nua}. In particular, for the
  present calculation,  no derivative coupling has been considered.
Then, the
properly symmetrized BSA of  the nucleon is approximated as follows
\begin{equation}
\Phi^\sigma(k_1,k_2,k_3,P_N) 
=\Lambda(k_1,k_2,k_3)~{\mathcal U}^\sigma(k_1,k_2,k_3,P_N)
 \label{ampli1}
\end{equation}
where $\Lambda(k_1,k_2,k_3)$ describes  
 the symmetric momentum dependence of the
vertex function upon the quark momenta, and ${\mathcal U}^\sigma $is given by
 \be{\mathcal U}^\sigma=
\left [ {\mathcal S}(123)+{\mathcal S}(312)+{\mathcal S}(321) 
 \right ] ~ \chi_{\tau_N} ~ U_N^\sigma\ee with $\chi_{\tau_N}$  the 
 nucleon isospin state and 
 \be {\mathcal S}(ij\ell)=\imath~\left [S(k_i)~ \tau_y ~ 
 \gamma^5 ~ S_C(k_j)C \right ]\otimes
 S(k_\ell)~~~~~~.\label{sym}\ee 
In Eq. (\ref{sym}), $C$ is the charge conjugation operator,  $S_C(k)= C ~S^T(k)~C^{-1}$ and 
the symbol $\otimes$  keeps separated the matrices
acting on the quark pair  from the ones 
acting on the quark-nucleon system.
Note that the  
symmetry in the quark pair variables reduces the number 
of possible terms from 6 to 3.

 The  quark-photon vertex ${\mathcal I}^\mu(k,q)=~
{\mathcal I}^{\mu}_{IS} +\tau_z ~ {\mathcal I}^{\mu}_{IV}$  has both isoscalar 
and isovector contributions. In turn,  each term, ${\mathcal I}^{\mu}_{i}$ (with $i= IS, IV$), is the
sum of  
i) a purely valence (V) contribution, ${\mathcal I}^\mu_{i,V}(k,q)$,
 which is present in the SL region only (see below) and  ii) 
  a nonvalence (NV) contribution,
 ${\mathcal I}^\mu_{i,NV}(k,q)$, corresponding to the $q\bar q$-pair
production (Z-diagram). Moreover,  
${\mathcal I}^\mu_{i,NV}(k,q)$ is 
 composed of a  point-like bare term and of a VMD term (cf 
 \cite{iach}).
 Summarizing one has
\be
 {\mathcal I}^\mu_{i,V}(k,q) = {{\mathcal N}_{i}}~\theta(P^+_N-k^+)~\theta(k^+)
 \gamma^\mu
 \nonu {\mathcal I}^\mu_{i, NV}(k,q) =\theta({q}^+ + k^+)~
\theta(-k^+)~\left [{Z_B ~{\mathcal N}_{i}} ~\gamma^\mu + Z^i_{VM} \Gamma^\mu_{VMD}(k,q,i)\right]
\label{vert} 
\ee
 where ${\mathcal N}_{IS}=1/6$, ${\mathcal N}_{IV}=1/2$. 
 The constants  $Z_B$, $ Z^{IV}_{VM}$ and $Z^{IS}_{VM}$ are  unknown weights 
 for the pair-production contributions, to be determined from the phenomenological
analysis of the experimental data. In principle, we should expect only one renormalization factor,
but in the actual   fitting procedure, we have taken $Z^{IS}_{VM}\ne
 Z^{IV}_{VM}=Z_B$, given the lower degree of knowledge of VM isoscalar sector. 
 We anticipate that, from our fitting procedure, the
 deviation from the equality is $\sim ~10$\%. As  in the case of  the pion \cite{DFPS},
the bare term $\gamma^\mu$ fulfills the current conservation 
in a covariant
model \cite{MFPPS2}.
For the VMD term,   we extended the microscopic model of Ref. 
 \cite{DFPS}, by  including the isoscalar mesons 
 and  by making the VM vertex trivially
 transverse  to $q^\mu$ \cite{MFPPS2} (this means $q \cdot\Gamma_{VMD}
=0$). The
 same  VM mass spectrum,  em  decay constants and total decay widths
 of Ref. \cite{DFPS} have been used for the isovector part of the VMD term. 
 As to the isoscalar term,  for $i=1,2,3$,  VM masses  and  the corresponding total decay widths
   have been  
 taken from PDG \cite{PDG}, while for $i>3$ we have calculated   the masses by using
  the mass operator 
 of Ref. \cite{FPZ02}, with an interaction parameter
 $w_{IS}= w_{IV} -0.27~GeV^2$ ($w_{IV}=1.556~GeV^2$), in order to follow the
  Anisovitch-Iachello law (see,
 e.g., \cite{DFPS} for the isovector case), and  we have adopted the same   
 total decay width
  $\Gamma_{n}^i=.150 ~GeV$ as we had for the isovector case.  The em decay
  constants, 
$\Gamma^i_{e^+ e^-}$, necessary for determining $\Gamma^\mu_{VMD}(k,q,i)$, have
been calculated with the model of Ref. \cite{FPZ02}, and  agree,  
within the  errors, with the corresponding  experimental
values of the known IV and IS  vector mesons \cite{PDG}. 
Finally, we considered up to 20 mesons 
 for  achieving convergence  at high $|q^2|$.

 Following  the pion case \cite{DFPS}, 
  the four-dimensional integrations 
  on {$k_1$} and  {$k_2$} in Eq. (\ref{spc1}) are regularized by assuming a  
  suitable fall-off of the momentum component of the BSA.
  Furthermore, in the integrations on {$k_1^-$} and {$k_2^-$}
  we consider only  the poles of $S(k_i)$, 
  namely we disregard the analytic structure of
   $\Lambda(k_1,k_2,k_3)$ and of the momentum components 
 of the VM amplitudes, present in $\Gamma^\mu_{VMD}(k,q,i)$ (see \cite{DFPS}), 
 since it affects  
   Fock sectors beyond the ones implicit in Figs. \ref{slfey} and 
   \ref{tlfey}. Then the covariance is only approximate (see Ref. \cite{DFPS} for a
   quantitative discussion in the pion case).

For the sake of concreteness, let us show the formal expression  of 
 the microscopic current
$I^{\mu}(q^2)$ (whose matrix elements are given in Eq. (\ref{spc1})) in the SL
region. It becomes
 the sum of two contributions: i) a purely valence (or triangle) contribution, 
$I_V^{\mu}(SL,q^2)$ (Fig. \ref{slfey},  diagram (a)),
where both the nucleon vertexes have two quarks on their $k^-$-shell 
($k^- = k^-_{on} = (|{\bf k}_{\perp}|^2 + m^2)/k^+ $) and the quark 
variables are in the valence region ($P^+_N\ge k^+_i\ge 0$); ii) a nonvalence (pair-production or Z-diagram) contribution,  
$I_{{NV}}^{\mu}(SL,q^2)$,
where the initial nucleon vertex has a quark  outside the valence 
range ($k^+_3<0$, see Fig. \ref{slfey}, diagram  (b))
  viz
\be
{ I}^{\mu}_{V}(SL,q^2)=-{3 ~ N_c  \over 2(2 \pi)^6}\int^{P_N^+}_0  {dk^+_1 \over k^+_1} 
\int^{P^+_N - k^+_1}_0   {dk^+_2 \over k^+_2} \int 
  { d{\bf k}_{1 \perp}~d{\bf k}_{2 \perp} \over k^+_3~
 k^{\prime +}_3}~ \Psi^*_N(\tilde k_1,\tilde k_2,{P'}_N)~\times \nonu \Psi_N(\tilde k_1,\tilde k_2,P_N)
  \left. {\mathcal F}^\mu
\right|_{(k^-_{1on},k^-_{2on})} 
\label{ISL}
\\ &&
{ I}^{\mu}_{{NV}}(SL,q^2)={3 ~ N_c  \over 2(2 \pi)^6}
 \int^{{P}_N^{'+}}_0 {dk^+_1 \over k^+_1} 
\int^{{P}_N^{'+} - k^+_1}_{P_N^+ - k^+_1}  {dk^+_2 \over k^+_2}  \int
  {d{\bf k}_{1 \perp} ~d{\bf k}_{2 \perp} \over k^+_3~
 k^{\prime +}_3} 
 \theta(k_2^+)~\times \nonu \Psi^*_N(\tilde k_1,\tilde k_2,{P'}_N)~
  {\left.\left \{\Lambda(k_1,k_2,k_3)~ 
  {\mathcal F}^\mu \right \}\right|_{(k^-_{1on},k'^-_{3on})} \over \left [q^- -
  {k'}^-_{3on} + k^-_{3on} \right ]}  
\label{IISL}
\ee
 where $\tilde k_i\equiv\{k^+_i,{\bf
 k}_{i\perp}\}$ is the light-front momentum   and $k'^-_{3on}=(P'_{ N} -k_1-k_2)^-_{on}$. The quantity 
 ${\mathcal F}^\mu$ is a
 $4\times 4$ matrix  (see \cite{nua} and \cite{MFPPS2}) constructed from 
 ${\mathcal
 U}^\sigma$ and ${\mathcal I}^\mu_{i}(k,q)$ (Eq. \ref{vert}),  given by 
 \be
 {\mathcal F}^\mu =  
  (\psla k'_3 +m)~{\mathcal I}^{\mu}_N
 (\psla k_3 +m)~Tr~{\mathcal K}(2,1) \nonu 
 +(\psla k'_3 +m)~{\mathcal I}^{\mu}_N 
 (\psla k_3 +m){\mathcal K}(2,1)+{\mathcal K}(2,1)(\psla k'_3 +m)~{\mathcal I}^{\mu}_N 
 (\psla k_3 +m)
 \nonu 
 -{\mathcal K}(1,3)~\gamma^5[{\mathcal I}^{\mu}_{IS} - \tau_N ~
  {\mathcal I}^{\mu}_{IV}]
 \gamma^5(\psla k'_3 +m)(\psla k_2 +m)\nonu 
 + 2 (\psla k_2 +m)~Tr[(\psla k'_3 +m){\mathcal I}^{\mu}_{IS}
 {\mathcal K}(3,1)]
 \label{tracce}
\ee
where ${\mathcal I}^{\mu}_N={\mathcal I}^{\mu}_{IS} + \tau_N ~
  {\mathcal I}^{\mu}_{IV}$ and  
  ${\mathcal K}(i,j)=(\psla k_i +m)(\psla k_j +m)$.

 In Eqs. (\ref{ISL}, \ref{IISL}) the momentum dependence of the vertex functions 
 in the valence range ($ P^+_N\ge k^+_i \ge 0$) is  
 expressed through a light-front wave function, $\Psi_N$, 
 which is  
a PQCD inspired wave function $a ~ la$ Brodsky-Lepage 
(see, e.g., \cite{brodsky}), described in terms of 
the squared free mass of the three-quark system
$M^2_{0N}(k_1,k_2,k_3)=P^+_N~\sum _ik^-_{i, on}$, i.e.
\be
~P^+_N~
{\left.\Lambda(k_1,k_2,k_3)\right|_{(k^-_{1on},k^-_{2on})}
\over M^2_{ N} - M^2_{0N}(k_1,k_2,k_3)} = 
\Psi_N(\tilde k_1,\tilde k_2,P_N)=\nonu=
{\mathcal {N}}~
{~P^+_N~(9~m^2)^{7/2}
 \over (\xi_1\xi_2\xi_3)^{p}~\left[\beta^2 + M^2_{0N}(k_1,k_2,k_3)\right]^{7/2}}
 \label{VWF}
\ee
where 
$\xi_i = k^+_i/P^+_N$ and ${\mathcal {N}}$ is a normalization constant, 
obtained from the plus component of
the proton current at $Q^2=0$, i.e. from the proton charge normalization. 
In Eq. (\ref{VWF}) the power $7/2$ 
and the parameter $p = 0.13$, which controls the end-point behavior and 
affects the FF 
 mainly through the Z-diagram,  allow one
to obtain an asymptotic decrease of the valence contribution faster than 
the dipole
$G_D(|q^2|)= [1 + |q^2|/(0.71 (GeV/c)^2)]^{-2}$.
Since the Z-diagram gives no contribution to the nucleon magnetic moments,
the parameter $\beta = 0.645~GeV$  in Eq. (\ref{VWF})
can be  directly fixed through a fit to the experimental values,
obtaining  $\mu_p^{th} =  2.87\pm 0.02$ ($\mu_p^{exp}= 2.793$)  and
$\mu_n^{th} =  -1.85\pm 0.02$ ($\mu_n^{exp}= -1.913$). 
Theoretical uncertainties come from the Montecarlo integration of (\ref{ISL}). 

In Eq. (\ref{IISL}), the vertex function
$\left.\Lambda(k_1,k_2,k_3)\right|_{(k^-_{1on},k^{\prime -}_{3on})}$ 
describes a $qq \bar q$ system since $k^+_3<0$, and therefore 
it cannot be approximated as the one in the valence region.
 It turns out \cite{MFPPS2} that this NV 
vertex leads to  a contribution to the nucleon FF to be interpreted
as a transition from $|qqq\rangle$ to $|qqq q\bar q\rangle$ 
Fock components of the final nucleon. In the present calculation, an 
 Ansatz,
${\Lambda}^{SL}_{NV}=\left.\Lambda(k_1,k_2,k_3)\right|_{(k^-_{1on},k^{\prime -}
_{3on})}$,
in terms of invariants as the squared free mass, $M^2_0(1,2)$, 
of the quark pair propagating from the initial nucleon
toward the final one, and the squared free mass of the system $N+ \bar 3$, 
has been adopted (cf diagram (b) in Fig. \ref{slfey})
\begin{equation}
  {\Lambda}^{SL}_{NV}=
  [g_{1,2}]^{2}[g_{N,\bar 3}]^{7/2-2}\left [{k_{12}^+
 \over P^{\prime +}_N }\right ] 
  \left [ P^{\prime +}_N \over k_{\overline {3}}^+ \right ]^r 
 \left [ P^{ +}_N \over k_{\overline {3}}^+ \right ]^r
\label{offsl}
\end{equation}
where $g_{A,B}=(m_A ~ m_B) / \left
[\beta^2+M^2_0(A,B)\right]$ and $k_{12}^+ = k_1^+ + k_2^+$.
The ratio  $k^+_{12}/P^{\prime +}_N$ enforces the
collinearity of the  spectator-quark pair and the final nucleon, while 
$[P^{\prime +}_N/k^+_{\bar 3}]^r ~[P^{+}_N/k^+_{\bar 3}]^r$ 
controls the end-point
behavior of the antiquark-leg attached to the nonvalence vertex (with a 
 chosen 
symmetrical form). 
The powers of $g_{12}$ and  $g_{N\bar 3}$ and the parameter $r=0.17$ allow 
one to obtain a dipole asymptotic behavior for the NV contribution.

In the TL region, where $q=P_N~+~ P_{\bar N}$,
$k_{12}+k_3=-P_{\bar N}$ and $k_{12}+k'_3=P_N$, after integrating on 
 $k_1^-$ and $k_2^-$, one obtains  two contributions, with a form similar to 
 Eq. (\ref{IISL}), but corresponding to diagrams (a)
 and (b) of Fig. \ref{tlfey}.
 In both  contributions  valence and 
NV nucleon vertexes  are present, as a result of   a  
transition between the
$ |q q q, \bar q  \bar q \bar q\rangle$ hadronic component of the photon
 state and the  $N\bar N$ final state. The nucleon NV vertex is approximated 
 by an Ansatz, ${\Lambda}^{TL}_{NV}$,
analogous to Eq. (\ref{offsl}), built up with the corresponding invariants, 
e.g., in the contribution (a) of Fig.  \ref{tlfey} one has
  \begin{equation}
    {\Lambda}^{TL}_{NV}= 2
  [g_{ \bar 1,\bar 2}]^{2}[g_{ N, \overline {1 2}}]^{7/2-2}    
 \left [{- k_{12}^+ \over P^+_{\bar N} }\right ] 
 \left [P^+_{\bar N} \over {k'}^+_3 \right ]^r 
 \left [P^+_{ N }\over {k'}^+_3 \right ]^r
\label{offtl}
   \end{equation}
where the factor  $2$ counts the possible patterns for gluon emissions.  
In $g_{ N, \overline {1 2}}$, we put in the  normalization factor    
$m_{\overline {1 2}}=0.500~ GeV$\cite{Traini}, and in the denominator $M^2_0(N, \overline {1
2})=(P^+_N -k^+_{1 }-k^+_{2 })~(P^-_N -k^-_{1 on}-k^-_{2 on})-
|{\bf k}_{1 \perp}+{\bf k}_{2 \perp}|^2$

\begin{figure}
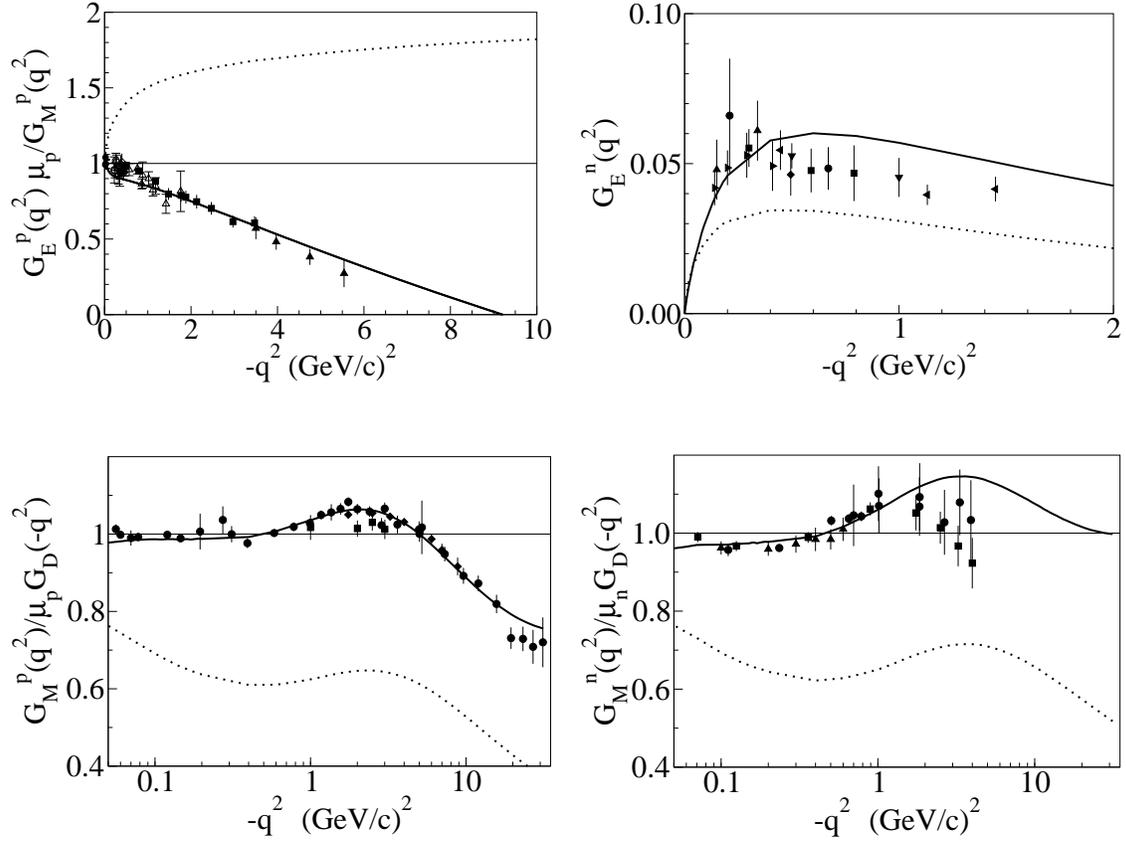

\parbox{7.2cm}{\includegraphics[width=7.2cm]{Ratio_plb.eps}}$~~~$
\parbox{7.2cm}{\includegraphics[width=7.2cm]{Gen_plb.eps}}

\vspace{0.9cm}
\parbox{7.2cm}{\includegraphics[width=7.2cm]{Gmp_plb.eps}}$~~~$
\parbox{7.2cm}{\includegraphics[width=7.2cm]{Gmn_plb.eps}}
\caption{\label{pdata}
Spacelike nucleon  form factors vs $-q^2$. 
 Solid lines: full calculation, i.e., sum of triangle plus pair-production terms. 
 Dotted
lines: triangle contribution only. Data from the compilations in \cite{Kees}. 
$G_D(|q^2|)= 
[1 +| q^2|/(0.71 (GeV/c)^2)]^{-2}$.}
\end{figure}

To determine the free parameters $Z_B$, $Z^{IS}_{VM}$, $p$ and $r$, 
a fitting procedure  has been performed in the SL region, including 
 proton data ($\mu_p~G^p_E/G^p_M$ and $G^p_M$)
with $Q^2\le 10 ~(GeV/c)^2$
and  neutron
 data ($G^n_E$ and $G^n_M$) with $Q^2\le 1 ~(GeV/c)^2$. 
We obtained  a value $\chi^2/datum=1.7$, with a very
nice description of the data, as shown in Fig. \ref{pdata}. From the
fitting procedure we have: i)
the ratio $Z_{VM}^{IS}/Z_{VM}^{IV}=1.12$, remarkably  close to one,
and ii)  
$Z_B=Z_{VM}^{IV}=2.283$.
In correspondence to the previous outcome, the proton
charge radius is $r_p=0.903\pm .004~fm$ ($r^{exp}_p= 0.895 \pm 0.018 $ \cite{Kees}) 
and $-dG^n_E(q^2)/dq^2=0.501\pm 0.002~(GeV/c)^{-2}$ (the exp. value is $0.512 
\pm ~0.013~(GeV/c)^{-2}$
\cite{Kees}). 

The same values for  $Z_B$, $Z_{VM}^{IS}$  and $r$
(see Eq. (\ref{offtl})) 
are adopted for calculating the
effective TL form factors, defined as follows, according to experimentalists
(see, e.g., \cite{BABAR}),
\be G_{eff}^{p(n)}(q^2) = \sqrt{(~|G_M^{p(n)}(q^2)|^2  -  \eta~|G_E^{p(n)}(q^2)|^2~)  
 /
(1 ~ - ~\eta) }\label{tlff} \ee 
 
The Z-diagram ({higher Fock components}) is essential for 
describing the  nucleon FF, 
in the adopted reference
frame  ($q^+ \ne 0$), as  in  the pion case \cite{DFPS}. 
In the SL region, it produces the striking
 feature of  a zero 
around $Q^2\sim 9.0~(GeV/c)^2$ for
  $G^p_E$. Notably,  retaining only three sets of data, 
  $G^p_M$, $G^n_E$ and $G^n_M$, 
in the fitting procedure, one gets again the
 cancellation  between triangle and 
 pair-production contributions
 to $G^p_E$,  and only tiny differences from  the results shown in Fig.
  \ref{pdata}. This means that, in our model, the falloff of $\mu_p~G^p_E/G^p_M$
  for $Q^2> 1.0~(GeV/c)^2$ is enforced by the other three sets of data.

\begin{figure}
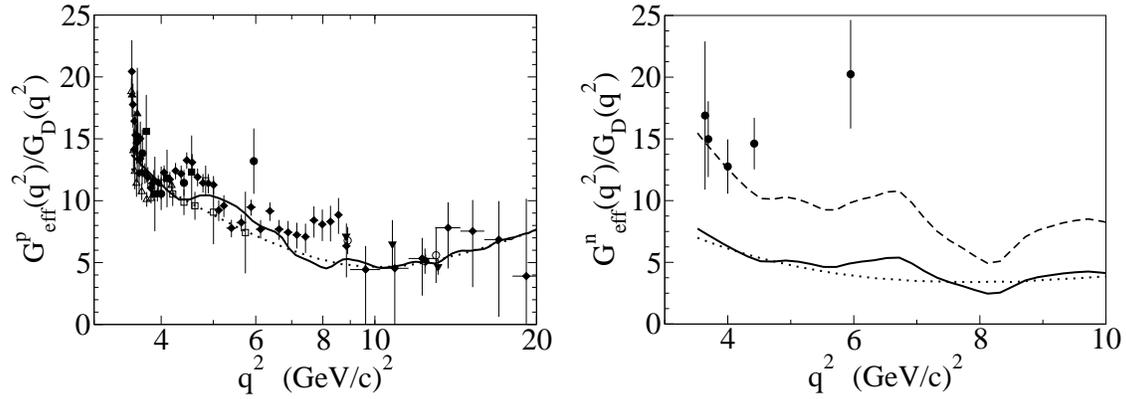

\parbox{7.2cm}{\includegraphics[width=7.2cm]{TLP_GD.eps}}$~~~$
\parbox{7.2cm}{\includegraphics[width=7.2cm]{TLN_GD.eps}}
\caption{\label{tldata} Nucleon effective form factors 
(see Eq. (\ref{tlff}) for the definition) in the
timelike region. Solid line: bare + VM. Dotted line: bare term. Left panel:   $G^p_{eff}(q^2)/G_D(q^2)$ vs $q^2$; data from \cite{BABAR}.
Right panel: $G^n_{eff}(q^2)/G_D(q^2)$ vs $q^2$; data from \cite{Fenice}. Dashed
line: solid line arbitrarily multiplied by  2.  }
\end{figure}

In the TL region, our calculations are parameter free, and give a fair 
description of the proton data,
 apart the peak
 at the threshold, which  is 
outside the present model due to the absence of the final state interaction. 
The TL proton data  clearly show a structure due to the resonances 
(see Fig. \ref{tldata}), allowing to gather more details on 
 the hadronic content of the photon
wave function.  
In particular,  the comparison with 
the most recent data \cite{BABAR}
 put in evidence that some strength is lacking in our model
 for $q^2\sim 4.5~ $ and $ \sim 8~ (GeV/c)^2$ (as for the pion
 \cite{DFPS}).
Finally,
available TL neutron data are not reproduced by the present model, but,
 even a constant  factor of 2 
could improve the
description (cf  Fig. \ref{tldata},  right panel).

Summarizing,  in a frame with $q^+\ne 0$ (that allows a unified
treatment of both the SL and TL regions),  our approach,
 with only four adjusted parameters ($Z_B=Z_{VM}^{IV}$,
$Z_{VM}^{IS}$, $p$ and $r$),  is able to describe  the 
 nucleon SL FF ($\chi^2/datum=1.7$) and to give predictions for  the  TL ones.
A complete analysis of the model dependence, as well as a detailed study of the  
momentum distributions of the valence nucleon vertex functions will be
 presented elsewhere, together with
 a study of  nucleon FF in the
  unphysical region ($0<~q^2< 4M^2_N$), which appears very challenging, 
  but  needs a
  non trivial inclusion of the $\bar N N$ interaction \cite{MFPPS2}.

\section*{Acknowledgments}
 This work was supported 
by the Brazilian agencies CNPq
and FAPESP and by the Italian MUR.

\newpage

\end{document}